\documentclass[a4paper]{article}
\pdfoutput=1
\usepackage{url}
\usepackage{graphicx}

\usepackage{floatpag} 

\usepackage{titlesec}
\titleformat*{\section}{\large\bfseries}
\titleformat*{\subsection}{\normalsize\bfseries}

\usepackage[round]{natbib}  
\setcitestyle{authoryear}

\usepackage[margin=1.7in]{geometry}
\usepackage{microtype}
\usepackage{setspace} 
\usepackage[space]{grffile} 
\let\origfigure\figure \let\endorigfigure\endfigure \renewenvironment{figure}[1][tbph]{ \origfigure[#1] \centering }{ \endorigfigure } 

\title{Where is the global corporate elite? \\A large-scale network study of local and nonlocal interlocking directorates}
\author{
Eelke M. Heemskerk, Frank W. Takes, \\ Javier Garcia-Bernardo and M. Jouke Huijzer \\ \url{{e.m.heemskerk,takes,garcia}@uva.nl} \vspace{2mm} \\
CORPNET, University of Amsterdam \\ 
\url{http://corpnet.uva.nl} }
\date{July 14, 2016}

\begin{document}

\maketitle

\begin{abstract}
Business elites reconfigure their locus of organization over time, from the city level, to the national level, and beyond. 
We ask what the current level of elite organization is and propose a novel theoretical and empirical approach to answer this question. 
Building on the universal distinction between local and nonlocal ties we use network analysis and community detection to dissect the global network of 
interlocking directorates among over five million firms. 
We find that elite orientation is indeed changing from the national to the transnational plane, but we register a considerable heterogeneity across different regions in the world. 
In some regions the business communities are organized along national borders, whereas in other areas the locus of organization is at the city level or international level. 
London dominates the global corporate elite network. 
Our findings underscore that the study of corporate elites requires an approach that is sensitive to levels of organization that go beyond the confines of nation states.
\\


\end{abstract}

\newpage

\section{Why we should \emph{not} study transnational corporate elites}

With the growing internationalization of business during the second half of the twentieth century, critical scholars increasingly saw a development where the elites who own and control corporations cease to be organized or divided along national lines. 
Instead, they expect the formation of a `transnational capitalist class' \cite[pp. 324]{burris2012search}. 
They predicted that the national identities of corporate elites would be replaced by a common transnational identity with a shared sense of economic interests and an enhanced capacity for unified political action~\citep{robinson2004theory,sklair2001transnational}. 
Robinson, for instance, holds in his Theory of Global Capitalism that the dynamics that caused the formation of national capitalist classes, are now unfolding on a transnational scale. 
After the hegemonic periods of respectively Holland, Britain and the US, ``the baton is being passed to an emerging transnational configuration'' and ``we are witness to an emerging transnational hegemony in a process that is contested and far from finished, the emergence of a new historic block that is global in scope and based on the hegemony of transnational capital'' \cite[pp. 77]{robinson2004theory}. 
Robinson expects that the main contradictions that emerge in the process of globalization will be between the nationally oriented bourgeoisie and the elites that have an interest beyond the nation states boundaries.

Unfortunately, much of the literature on the emergence of a transnational capitalist class drew on a mixture of anecdotal evidence and theoretically informed speculation, as is rightly pointed out by \cite[pp. 324]{burris2012search}. 
Notable exceptions exist however. 
\citet{fennema1982international}, for instance, was the first to empirically analyze the transnational network of interlocking directorates – directors that sit on the boards of two companies and thus form a connection between the two firms. 
He found that by 1970 there was an Atlantic component of interlocking directorates, which had national clusters connected by international interlocks. 
By 1976 this network had become much more transnational; the number of international interlocks in the Atlantic component had increased by 50 percent. 
However, the transnational network of board interlocks was primarily established between neighboring countries in Europe, crossed to a lesser extent the Atlantic, and hardly connected upcoming economic regions in Asia or the rest of the world~\citep{fennema1982international}. 
Board interlocks proved a helpful element in the empirical studies on social corporate elite organization. 
Scholars have argued that board interlocks serve to reduce economic uncertainty and secure resources from banks and suppliers. 
Others see them as a mean to maintain class cohesion~\citep{mills1956power,zeitlin1974orporate}, 
integrate new elites and facilitate class-wide political action~\citep{domhoff1970}.
Other perspectives hold that board interlocks together form a communication system and provide a `business scan'~\citep{useem1984inner} allowing for the spread of business strategies and information about (future) economic trends and challenges (see \cite{davis1997corporate}, or \cite{mizruchi1996interlocks} for a review). 
Of course, we should not reduce the entire (transnational) corporate elite community (and its various sources of integration) to the practice of interlocking corporate directorships. For sure, there are many key actors in the business community outside the set of corporate board members, such as lawyers, accountants, but also fund managers and perhaps even regulators. Yet, whatever their purpose, board interlocks are present in practically every country and, as such, are studied being a meaningful indicator of elite social organization~\citep{david2014power,scott1991networks}. 

Most empirical work on business elite networks in the twentieth century looked exclusively at connections within particular countries. 
The transnational corporate elite was debated and theorized, but, aside from the work of \cite{fennema1982international} and \cite{fennema1985transnational}, 
 empirical research remained scarce. In the twenty-first century, however, increased attention was paid to the transnational network. 
A follow up study on the pioneering work of~\cite{fennema1982international} by \cite{carroll2002there} 
 on the transnational network found that twenty years after 1976, the transnational network of interlocking directorates remained remarkably stable. 
Although it was coined the age of globalization, by the mid-1990s the network of transnational corporate elites still comprised mainly corporate directors and executives from the leading capitalist countries in North America and Europe (see also \cite{kentor2004yes}). 
In the early years of the twenty-first century, things start to change. 
By the mid-2000s the top of the transnational network also connected to elites in countries such as India, Mexico, Singapore, Turkey, Brazil and China, leading \cite[pp. 233]{carroll2010making} 
 to conclude that a transnational capitalist class was ``in the making, but not (yet) made''. 
Most recent empirical work underscores the resilience of the transnational network in the wake of the decline of national network cohesion and point at the increasing relevance of these transnational ties~\citep{heemskerk2016global}. 
A transnational corporate elite may have finally arrived. 
 
A consistent but often overlooked finding is that the increase in transnational corporate networks typically does not connect far away regions in the world, but rather integrates business elites that are relatively nearby, such as in Europe, or North America~\citep{carroll2010making,carroll2010constituting,fennema1985transnational} 
And this may have important repercussions for those who expect a transnational corporate elite identity to emerge. 
After all it is fair to say that a board interlock between a German and a Dutch, an USA and a Canadian, or a Venezuelan and a Colombian firm is qualitatively different from ties that span different regions of the globe, and connect for instance firms in the USA with China, Spain with India, or Italy with Brazil. 
This begs the question whether these nearby ties, although they connect the boards of firms domiciled in different countries, are indeed the building blocks of global elite networks. This question is part of the more general problem of theoretical and methodological nationalism. 
Theoretically, `transnational' is always defined in relation to the `national'. 
Empirically, we often only have data available that allows us to study and compare network patterns between countries. 
If our theoretical concepts are based on countries, and our empirical data is organized in countries, it may come as no surprise that we find that countries are important. For the literature on corporate elites this distinction in national and international has led to the search for elite networks that stand free from national business communities as a hallmark sign for the transnational corporate elite. 
But by assuming that global corporate elites emerge when they transcend national borders, one fails to acknowledge that corporate elite networks are often delineated by boundaries that not necessarily coincide with that of the nation state. 
Historical, cultural, geographical and political factors determine how and where the global corporate elites are organizing and whether the nation state is a meaningful spatial category to depart from. 
We should therefore make an effort to develop theoretical and empirical approaches that leave behind the a priori focus on the nation state as the most relevant spatial category for corporate elite organization.

\section{Local and nonlocal corporate elite networks}





\subsection{The reorganization of the local}

In order to accommodate our effort to come up with an approach that does not container corporate networks a priori within the boundaries of the nation state, we turn to~\cite{kono1998lost}, 
 who introduced the useful distinction between local and nonlocal ties. This distinction is helpful because it allows the analysis of the structure of networks of interlocking directorates without presupposing that the nation state is the most important level of elite organization. The dominant approach in the contemporary literature is to think about national corporate elite networks as the local, and consider transnational ties that cross national borders as the nonlocal. But the increase in corporate elite network cohesion across Europe mentioned above points at alternative reading: it may well be the case that what is local for corporate elites is what is changing. And this would not be the first time either. From detailed country studies, we know that in the first half of the twentieth century, corporate elites where typically organized around cities~\citep{dooley1969interlocking,levine1972sphere,allen1978economic}. 
In the USA, the Boston banking elite was different from the one in New York~\citep{pak2013gentlemen,mizruchi2013fracturing}
Even in a small country as the Netherlands, the pre-World War II corporate elite was organized in the distinct elite communities of Amsterdam, Rotterdam, The Hague (with ties to the Dutch colonies) and the textile industry around Enschede~\citep{heemskerk2007decline}. 
There is ample evidence of how these local elites organized their cohesion through upper class social interactions. 

Drawing on urban growth machine theory of~\cite{molotch1976city}, \cite{kono1998lost} argue that well into the twentieth century, the growth and business opportunities of most firms were predominantly a local affair, which required effective coordination by public and private actors within cities. 
To achieve this, corporate elites and state officials knit together in local growth coalitions in order to shape public policy that fosters local growth and as such advance their interests. 
If successful, some of the larger firms were able to expand their activities beyond the local growth coalition. 
This led to nonlocal ties between cities; sometimes in the same country, sometimes crossing national borders as well.
According to \cite{kono1998lost}, local ties and nonlocal ties are differently motivated and established by different sorts of firms. 
Firms that depend on local growth interlock primarily with local firms. 
Conversely, firms with suppliers and production outside the local region are more likely to interlock with geographically more distant firms.

This organization of elite cohesion at the local city level changed over the course of the twentieth century when corporate elites gradually integrated with other established networks within the boundaries of the nation state. 
In the second half of the twentieth century we see how these separate local elite communities that were primarily centered around particular cities, gradually merge together in national corporate elites. 
This transformation is perhaps best exemplified in the mergers of banks that used to serve the interest of regional business communities and in the emergence of national networks of interlocking directorates~\citep{heemskerk2007decline,mizruchi2013fracturing,david2014power}. 
What was considered as local for the business elite changed, and with it their networks changed as well. 
By the last quarter of the twentieth century, social organization of corporate elites at the local level meant in most areas in the world social organization at the national level. 

And with the change of the local, the nonlocal changed as well. 
The oil crisis in the 1970s prompted the now national corporate elites to invest in new nonlocal (and hence transnational) ties with corporations outside of their national business communities (as pointed about above). 
And in line with the expectations of \cite{kono1998lost}, 
 firms that engaged in these nonlocal transnational interlocking were typically production firms with a global reach, such as the globally operating oil companies (Shell, BP, Gulf Oil and Imperial Oil), car producers (Ford, FIAT, General Motors) and steel firms (Rio Tinto, British Steel)~\citep{heemskerk2016global}. 

The local, thus, varies in scope over time and space. 
In this article, we do not hold a predefined conception of what is local and what is nonlocal, as we accept that what is local may differ across various areas of the world. That business communities knit together and reconfigure the local in one place, does not mean that the same processes are taking place elsewhere. 
That some subnational business communities transformed into national business communities does not mean that all business communities became national in scope, nor that this transformation always took place along national borders. 
Consequently, we maintain that what is local is to be inferred from the empirical, rather than forced upon the empirical.

\subsection{What is local in a globalized world?}

There are good reasons to expect that for the contemporary corporate elite, again a reconfiguration of the local is taking place in many areas in the world. At least three empirical observations point into this direction. 
First, as mentioned above, the increase in transnational board interlocks in the twenty-first century~\citep{carroll2002there,kentor2004yes}. 
Here it is important to mention that transnational interlocking predominantly takes place within certain regions such as Europe and North America. 
Second, while transnational board interlocks used to be created by a small elite group of corporate directors, it has now become much more common practice among corporate elites.
A study of the European network of board interlocks in 2005 and 2010 showed that the hard core of European corporate directors who created at least four Pan-European interlocks remains stable in size (resp. 16 and 17 directors), but their contributions to the overall network density drops from 46 percent in 2005 to 25 percent by 2010~\cite[pp. 92]{heemskerk2013community}.
In similar vein, for the global network of board interlocks between a group of top 176 global firms we see that the contributions of the big linkers (at least four positions) drops from 31\% in 1976 to 20\% in 2006 and 12\% by 2013. The role of single linkers, directors with two positions, increases at the same time from 45\% in 1976 to 57\% by 2013~\cite[pp. 78]{heemskerk2016global}. 
This suggests that it becomes more common for corporate directors to engage in transnational board interlocks, and that the transnational network is no longer the domain of a few super connectors. Third, a recent study of the global network of interlocking directorates among the largest one million firms in the world showed that although we see increasing density of the European corporate network over the past decades, the underlying network structure remains in fact very regionalized. Because this study was not restricted to the top 300 or 500 largest firms it provided a much more fine-grained perspective on the social structure of Europe's corporate elite. Where \cite{carroll2002there} 
 concluded that by 1996 the transnational corporate elite network was still best described as a superstructure that rested upon rather resilient national bases, \cite[pp. 112]{heemskerk2016corporate} find a ``multilevel structure where, in between the national and the transnational, discernible regional clusters play a fundamental role in the network architecture''.
 
In what follows we study the patterns of local and nonlocal corporate elite organization. In today's world, what is considered local may very well cross national borders. And conversely, there may be nonlocal ties within one country as well. Imagine emerging markets, where some parts of the business elite try to integrate with the western elites and others are more oriented to their respective states. The truly global part of the corporate elite network is that which connects across business communities, disregarding its geographic position or scope. This raises the question of how to distinguish between local and nonlocal ties, without using a geographic definition such as national vs. international. In what follows we use a network analysis approach of community detection as a means of locating local and nonlocal ties based on the properties of the network structure itself.

\section{A large-scale network analytic approach to spatial elite organization}

\subsection{Community detection}

From a network perspective, communities are groups of nodes that share more connections with each other than with others outside the community. Communities are dense interconnected pockets of a network, where its members have a high degree of interconnectedness. We can define local ties as the set of ties that connect nodes that are all a member of the same community. Local ties are intra-community ties, regardless of whether this community is bounded by subnational, national or regional boundaries. And in similar vein, ties that connect points across different communities can be designated as nonlocal ties. Following this approach of local and nonlocal we do not assume that geographic distance or political dividing lines such as national borders are crucial for the formation of communities. But at the same time we do not preclude it either. This approach allows us to make the extent to which the corporate elite is organized along national borders an empirical question. It also allows for diversity in the (multi-level) organization of corporate elites from place to place. The local business community may at some places be a network centered around a particular city, whereas in other regions it is centered around a country or a transnational region. Hence, we reveal the geographic organization of the global corporate elite, and to do so we do not start with a geographically informed notion of local and nonlocal.

Previous studies typically focused on the networks of board interlocks between a few hundred of the largest corporations. Here we take a different approach and study a database that covers millions of firms across the globe. The benefit of this approach is that it releases us from the need to look at the most likely places of elite integration, and actually study the entire universe of interlocking directorates. In our analysis we study the network of board interlocks between a set of over 5 million connected firms across the globe. As we are interested in the geographic organization of the corporate elite, we consider the city locations of the head offices of all firms. This allows us to study how these locations are clustered in communities and which cities play a more or less important role in the creation of the nonlocal network ties. Yet, studying this number of board interlocks without delimiting our sample to a predetermined subset of firms also raises a question on whether all ties under study can be considered as elite-ties. A board interlock between two large firms whose operations span the globe supposedly serves different purposes than a tie between two small private firms connecting two nearby cities. From a global perspective, the latter tie can hardly be regarded an elite tie, while at a lower level of analysis the tie may still reinvigorate local growth coalitions. Rather than identifying the different characteristics of various corporate elite ties, our effort consists of an attempt to geographically locate the ties that span current local growth coalitions and are therefor most likely to accommodate  global or regional elite integration. 
Since this is the first time the global network of board interlocks has been studied in this scope we first describe and analyze its overall structure before we distill pattern of communities. Subsequently we investigate which cities are central in spanning the networks within and between communities. 

\subsection{Methods} 
\label{sec:datamethods}

To better understand our board interlock network, we use social network analysis techniques. First, we are interested in finding communities, a process commonly referred to as community detection. We choose to do so using modularity maximization. Modularity is a measure that indicates the quality of a division of the entire network into non-overlapping communities. A higher modularity value means that over the entire network, there are more connections between nodes in the same cluster than between nodes in different communities. Because the problem of finding the best division of a network into communities is NP-complete, community detection algorithms try to maximize the modularity value using heuristics and approximations. We use the common Louvain method~\citep{blondel2008}.
To assess the stability of the communities we find, we compared the results with some other modularity maximization algorithms that maximize the value using randomized methods. We observed no notable differences, apart from small artifacts of randomization. 

Second, centrality measures allow nodes to be ranked based on their structural position in the network. We use degree, eccentricity and betweenness centrality. The degree centrality measure determines for each node a centrality value based on the degree (number of connections) of that node, where a node with a higher number of connections is assumed to be more central. Degree centrality only considers the local neighborhood of a city as it counts the number of direct connections. It does not consider how a city is positioned in the network at large, for instance if it is in the core or in the periphery of the network. For this we turn to the second centrality measure, eccentricity. For each node in a network, eccentricity considers the shortest paths to all other nodes in that network. The longest of such paths gives the eccentricity score. For example, city that has an eccentricity of five can reach all other cities in at most five steps~\citep{takes2013computing}. 
	Third, betweenness centrality considers a node to be more central if it is part of a relatively larger number of shortest paths between all other node pairs, i.e. if it is a bridge between communities~\citep{brandes2008variants}. 
  
\subsection{Data}

The data source is the ORBIS database of Bureau van Dijk. This database contains information on over 200 million private and public companies across the globe. The ORBIS database has frequently been identified as one of the best sources of information for research on corporate networks~\citep{vitali2011network,compston2013network}. 
At the same time, we need to take into account the limitations of this data as well. It is sourced from various local information providers (such as national registers and chambers of commerce), and the quality of the data does differ with a bias towards better quality for richer countries. 
An exception to this trend is North America, whose data quality is moderate. Thus, our results may be downplaying the role of the United States in the global network
For a more elaborate discussion on the quality of the ORBIS data, see \cite{heemskerk2016corporate}. 
Notably, we do not limit ourselves to listed firms as that would exclude large parts of the economy. In this we follow common approach in the literature on global corporate board networks. 

\begin{figure}[!b]
	\includegraphics[width=\textwidth]{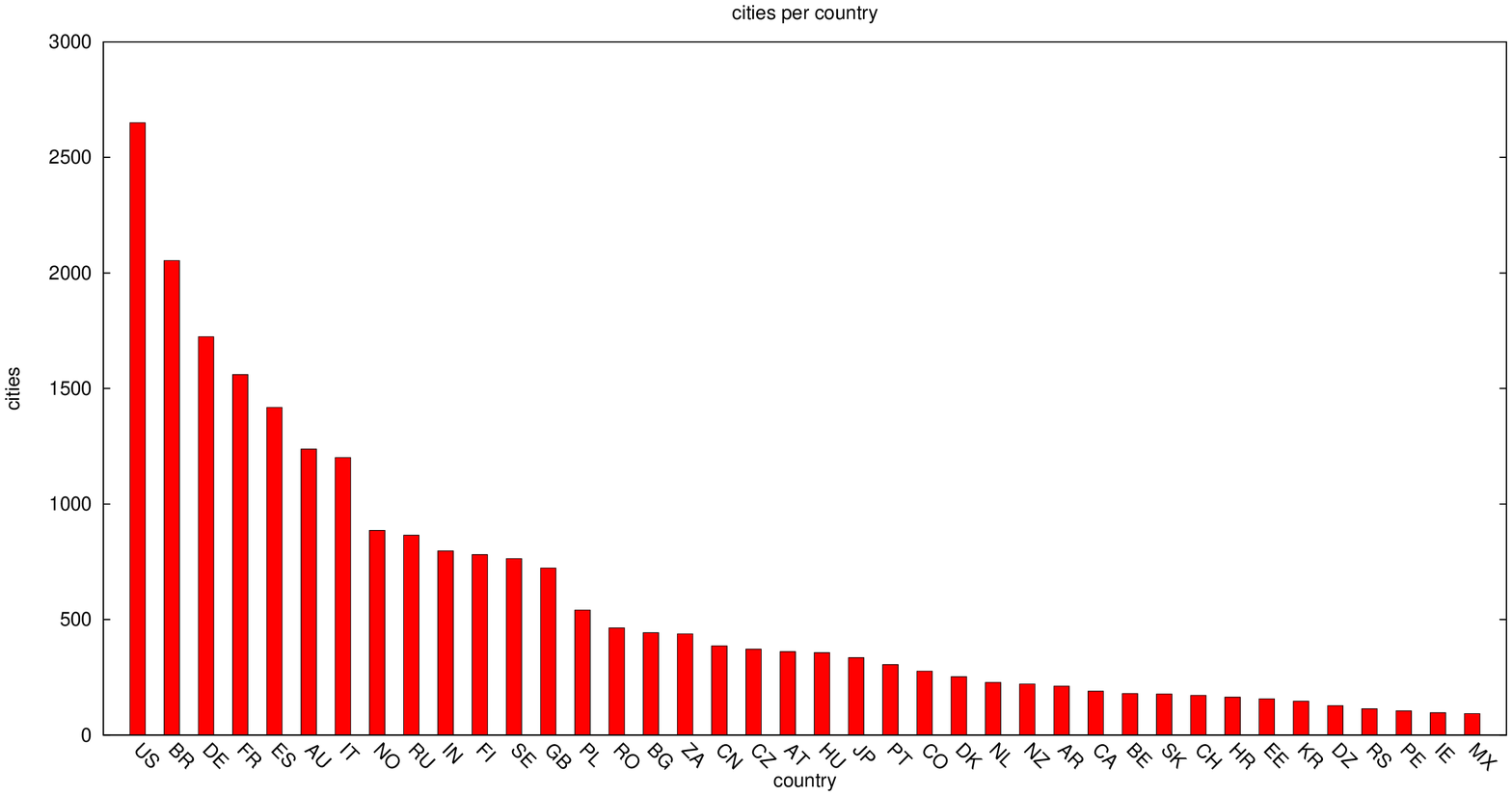}
	\caption{Division of cities over countries}
	\label{fig:countrydist}
\end{figure}

In selecting the firms and directors to construct the board interlock network, we proceeded as follows. 
In September 2015, we selected from the database all firms listed as `active' for which data on positions was available. 
For each firm, we acquired data on its positions, selecting only those positions listed as `current' and held by actual persons (and thus not by firms). 
We only selected positions listed as being the chief executive officer, highest executive, supervisory board, executive board, board of directors or member of a committee. 
This results in a set of 18,211,838 firms where 8,090,796 senior level directors hold positions such that an interlock is created between two firms. 
The cities of these firms are divided over countries as shown in Figure~\ref{fig:countrydist}.

An increasingly important problem when studying board interlock networks has to do with rise in complex administrative holding structures that firms create for fiscal and legal purposes. In the myriad of legal entities that together constitute a firm or bank, it becomes very difficult to pinpoint what the proper delineation of the firm is, and which board we should consider in our analysis. Shell companies and brass plate companies in tax havens only add to this confusion. As a result, there is a fair share of board interlocks in the network that are administrative ties that exist on paper only, rather than social ties where people actually meet and participate in meaningful board conversation and decision making. It is difficult to filter out these different types of ties~\citep{heemskerk2016corporate}. Here we have chosen a practical and effective approach to diminish the distorting effect of administrative ties. We eliminate the interlock ties that go hand in hand with an ownership relation of more than 50\% (as registered in ORBIS), as this is then likely a parent-subsidiary or holding structure. Similarly, we ignored directors with more than 100 positions, because it is unlikely that such a person facilitates `real' interlocks. 

The information on board positions is essentially a two-mode network of firms and officers linked through positions. 
From this we constructed a one-mode network in which the nodes represent firms and weighted edges represent shared senior level directors. 
The edge weight denotes the number of shared directors between the two firms the edge connects. 
In this firm-by-firm network, not all firms are connected to each other. 
In particular, the network contains multiple connected components. 
As is typical for these networks, there is one giant component. 
These 5,262,534 firms capture the majority of the interlocks. 
The second largest component in terms of size contains only 339 firms, and in fact 95\% of the smaller components are of size smaller than 20. 
The majority of economic activity is captured by the giant component, and for the remainder of this paper we will therefore focus on this component. 

So, we end up with a network of over 5.2 million firms that are connected by about 37 million board interlocks (the largest connected component in the network). 
For our analysis we consider the board interlocks between cities. 
To do so, we aggregate all firms that are domiciled in the same city in one node, and connect city nodes by means of a weighted edge indicating the number of interlocks between firms in these cities.
Appendix~\ref{sec:cityclustermethod} gives detailed information on how we aggregated the firms in city clusters.

\section{Empirical results}

\subsection{Network topology}

\begin{figure}[t]
	\includegraphics[width=0.6\textwidth]{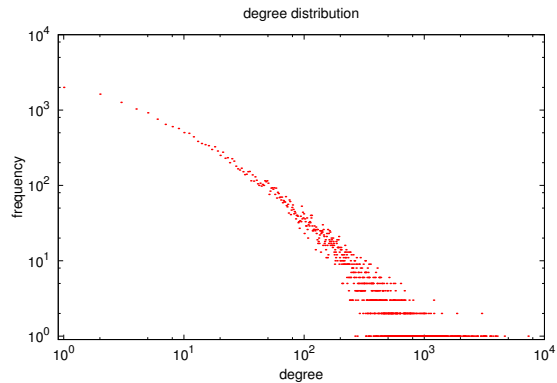}
	\caption{Degree distribution of the city network}
	\label{fig:degdist}
\end{figure}

We study the network of 24,747 cities, connected through 874,810 distinct ties (which together carry the 37 million board interlocks). Because the network is an aggregation of the connected component of the underlying firm-by-firm network, all cities are connected in one component as well. On average, each city is connected to 71 other cities. However, as displayed by Figure~\ref{fig:degdist}, there is a significant difference in the degree of connectivity of cities. A sizable group of cities have board interlock connections to no more than ten other cities, while a relatively small number of cities are super connectors, reaching out to over a 1,000 other cities. 
%
%
%
This power-law distribution signals the scale-free property of the network. 

\begin{figure}[b]
	\includegraphics[width=0.6\textwidth]{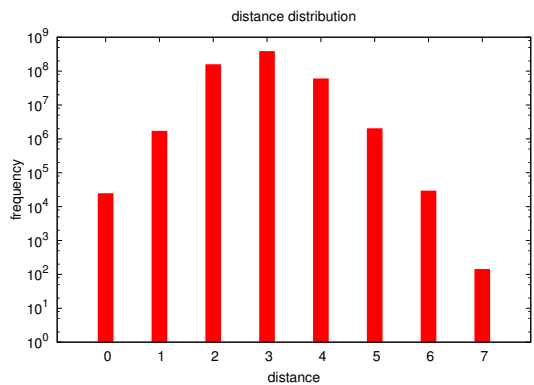}
	\caption{Distance distribution of the city network}
	\label{fig:distdist}
\end{figure}

How globally connected is the corporate elite? The average distance measures the average number of steps to be traversed to get from one node to another node. 
It turns out that cities in the global network of interlocking directorates are interconnected with an average distance of only 2.83. 
The distribution of average distances shown in Figure~\ref{fig:distdist} indicates that the global city network is relatively well connected through interlocking directorates; it is indeed a small-world network~\citep{kleinberg2000small}. 

A next step is to consider those cities that are most central in this global city network. 
Table~\ref{tab:eccdist} shows the distribution of the eccentricity values. 
On the rightmost end of the eccentricity spectrum there are the cities that are least well connected. 
The largest eccentricity score is 7, meaning that those cities are 7 steps away of at least one other city in the network. 
We can say these cities form the periphery. 
There are only 37 nodes in this periphery, which upon manual inspection appears to mainly consist of smaller businesses in rural areas of Russia and the United States. 
More interesting in Table~\ref{tab:eccdist}
 is the lowest value of eccentricity in the network. 
This value is called the radius, and the set of nodes with an eccentricity value equal to this this radius are in the center of the network. 
The cities in the center of the network have an eccentricity of 4, meaning that they can reach all other cities in at most four steps. 
This group consists of 409 cities: 1.65\% of the nodes in the city network. 

\begin{table}[t]
	\centering
	\caption{Eccentricity distribution of the city network} 
	\vspace{2mm}
	\label{tab:eccdist}
	\small
	\begin{tabular}{r|cccc}
		\hline
		\textbf{Eccentricity value}	& 4 & 5 & 6 & 7 \\
		\hline
		\textbf{Frequency} & 409 & 20782 & 3519 & 37 \\
		\hline
	\end{tabular}
\end{table}


\begin{figure}[!b]
	\includegraphics[width=\textwidth]{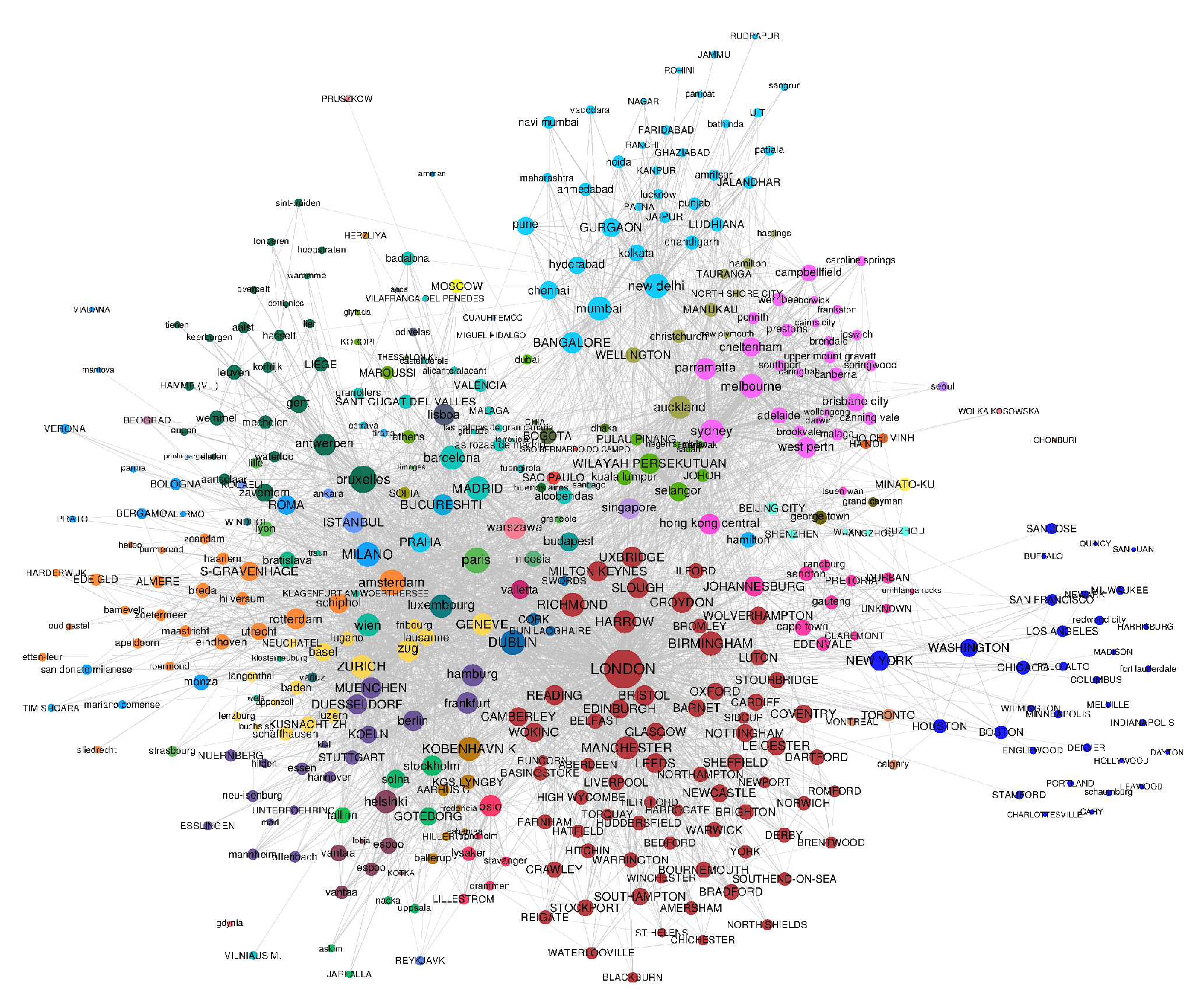} 
	\caption{Visualization of the center (eccentricity value 4) of the city network}
	\label{fig:center}
\end{figure}

Figure~\ref{fig:center} shows the network between this set of central cities. For visibility reasons, we have chosen to only show edges when two cities are connected by more than 20 board interlocks.  One observation clearly stands out: London is right in the center of this group of central cities. It is the most central city in the global board interlock network by both degree and betweenness measures (see Table~\ref{tab:centralities}). It is followed by high degree centrality cities Paris, Madrid, Brussels, Milano, Luxembourg and Barcelona. Sao Paulo and Moscow score among the top five cities ranked by betweenness centrality, since they connect their emerging markets to the global corporate network. Cities in the United Kingdom are well connected as well, not only to other UK cities but also abroad. In fact, UK cities connect more to cities abroad than within the UK. This effect is largely driven by London, which is extraordinary well connected. This overrepresentation of international city ties is typically a feature of small countries where businesses are concentrated in a few cities, such as Belgium, Switzerland and Austria. A final observation is that although there is a fair share of USA cities in this central core, they form somewhat of a peripheral cluster, primarily connected to the rest of the world through New York, Washington and Los Angeles. However, low data quality in the USA may downplay its role in the global network (Section~\ref{sec:datamethods}).

\begin{table}[t]
	\centering
	\caption{Top-25 most central cities in the city network} 
	\vspace{2mm}
	\label{tab:centralities}
	\small
	\begin{tabular}{ll}
		\hline
		\textbf{Degree centrality}	& \textbf{Betweenness centrality} \\
		\hline
		London	& London \\
		Paris	& Sao Paulo \\
		Madrid	& Paris \\
		Bruxelles	& Madrid \\
		Milano	& Moscow \\
		Luxembourg	& Melbourne \\
		Barcelona	& Sydney \\
		Wien	& Bruxelles \\
		Amsterdam	& Wien \\
		Dublin	& Barcelona \\
		Oslo	& Helsinki \\
		Helsinki	& Milano \\
		Sydney	& Oslo \\
		Zurich	& Bogota \\
		Kobenhavn	& Luxembourg \\
		Melbourne	& Rio De Janeiro \\
		Hamburg	& New Delhi \\
		Auckland	& Budapest \\
		Stockholm	& Auckland \\
		Zug	& Amsterdam \\
		Berlin	& Sofia \\
		Muenchen	& Washington \\
		New Delhi	& Praha \\
		Frankfurt	& Stockholm \\
		Praha	& Bucureshti \\
		\hline
	\end{tabular}
\end{table}

\subsection{Local communities in the global corporate elite}

Now we turn to the key objective of this paper and distinguish between local and nonlocal network ties. As explained above, we use the method of community detection and define the ties within communities as local, and those between communities as nonlocal. When we apply the Louvain modularity maximization method for community detection at the default resolution, all cities in the network are grouped in 14 separate communities. This division in communities has a modularity value of 0.83. Previous work has studied in depth how the community structure of global networks of interlocks is hierarchically nested~\citep{heemskerk2016corporate}. 
Here we choose to study the division in communities that generates the highest modularity value, that is, the partitioning in communities that optimizes the rule that nodes are better connected within a community than between communities. 

A fair share of the board interlocks occurs between firms within cities. This means that in the network, there are self-loops which `connect' cities with themselves. Some of these self-loops have considerable weight. If we include these self-loops when applying community detection, the communities we find are slightly biased towards different countries' capitals, and fewer communities appear. London has an extraordinary strong self-loop. If self-loops are taken into account when applying community detection, London is considered as a separate community. Without the integrative power of London the rest of the British business community falls apart in 19 separate additional communities. 

\begin{figure}[!t]
	\includegraphics[width=\textwidth]{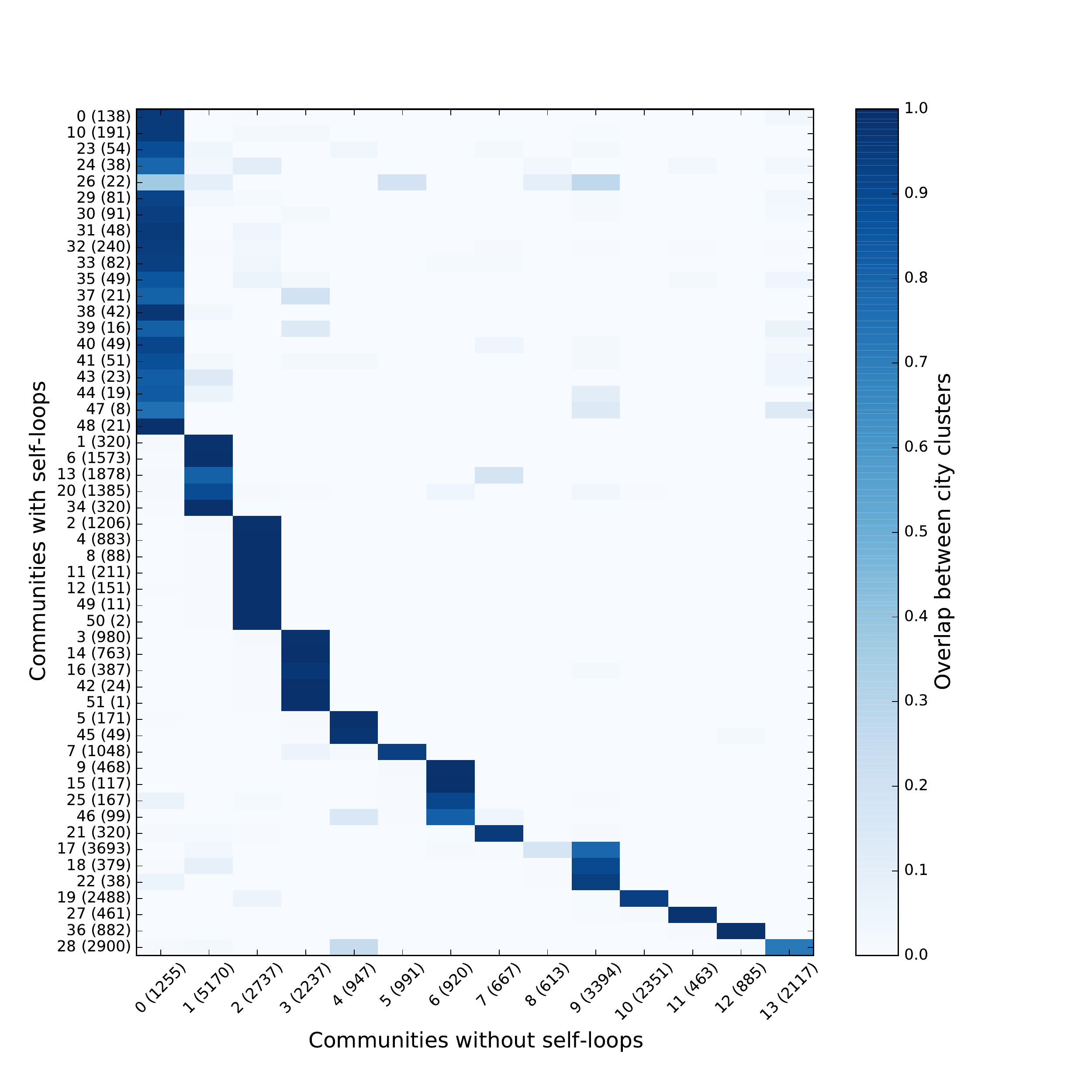} 
	\hspace{-12mm}
	\caption{Nested structure of communities with and without city self-loops}
	\label{fig:nestedself}	
\end{figure}

The cities in the UK in general and London in particular stand out in this respect. From the analysis with self-loops one can cautiously infer that the corporate communities are more sub-nationally structured than along national boundaries. Here we choose to exclude the self-loops for the community detection analysis because the large increase in communities makes our analysis of inter-community ties more difficult to interpret. Because of the hierarchically nested community structure, we do not throw away information but rather analyze the network at a higher level of aggregation. Figure~\ref{fig:nestedself} illustrates this by comparing how the communities with and without self-loops are hierarchically nested. The most leftward column shows for instance that the 20 separate British communities that are generated with self loops are all part of the British community in the analysis without self-loops. 

\begin{figure}[ph]
	\vspace{-35mm}
	\includegraphics[angle=-90,width=1.05\textwidth]{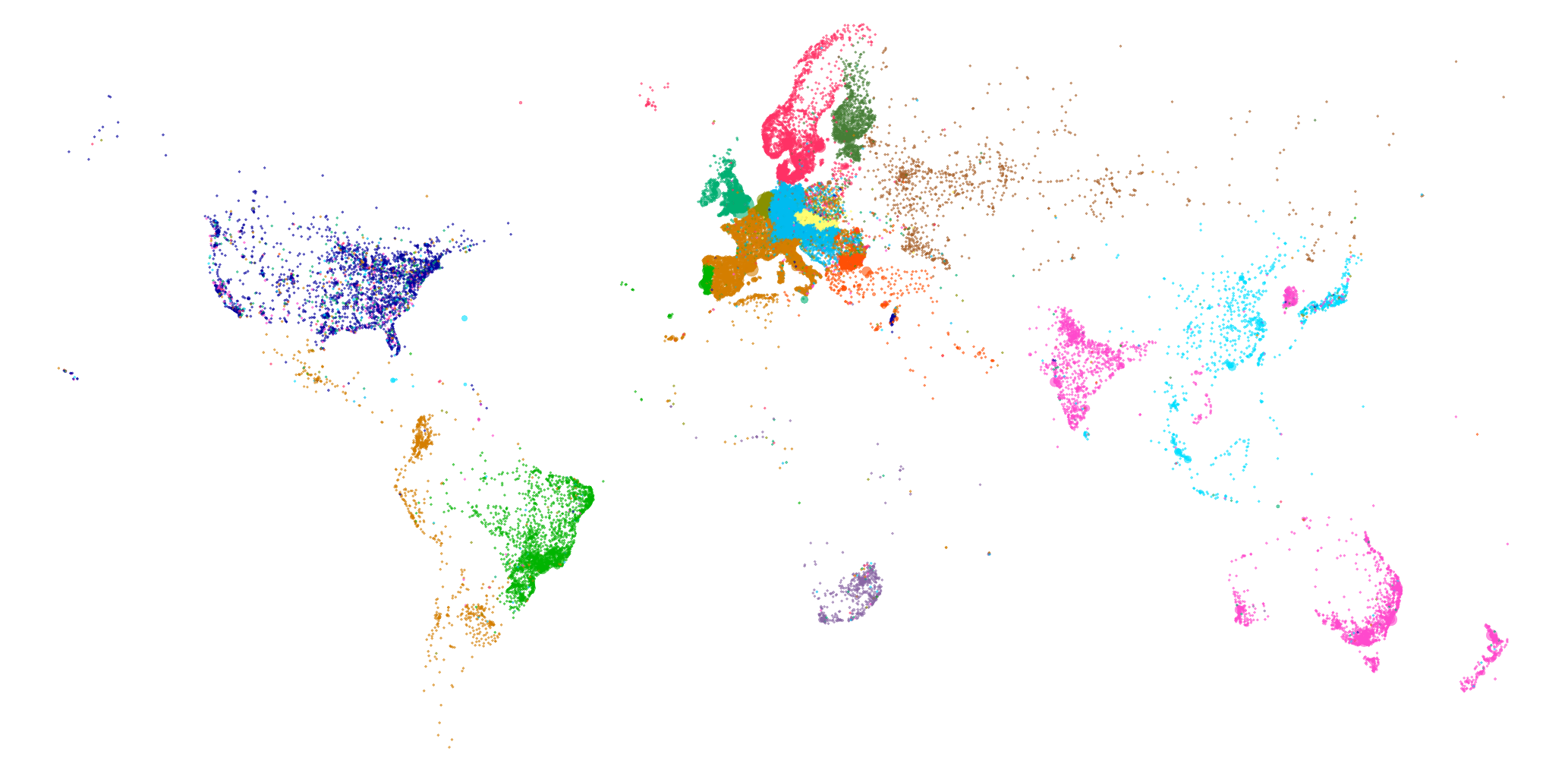} 
	\vspace{-5mm}	
	\caption{Community structure of the global network of interlocking directorates}
	\label{fig:global}
	\floatpagestyle{empty}
\end{figure}

\begin{figure}[tb]
	\includegraphics[width=\textwidth]{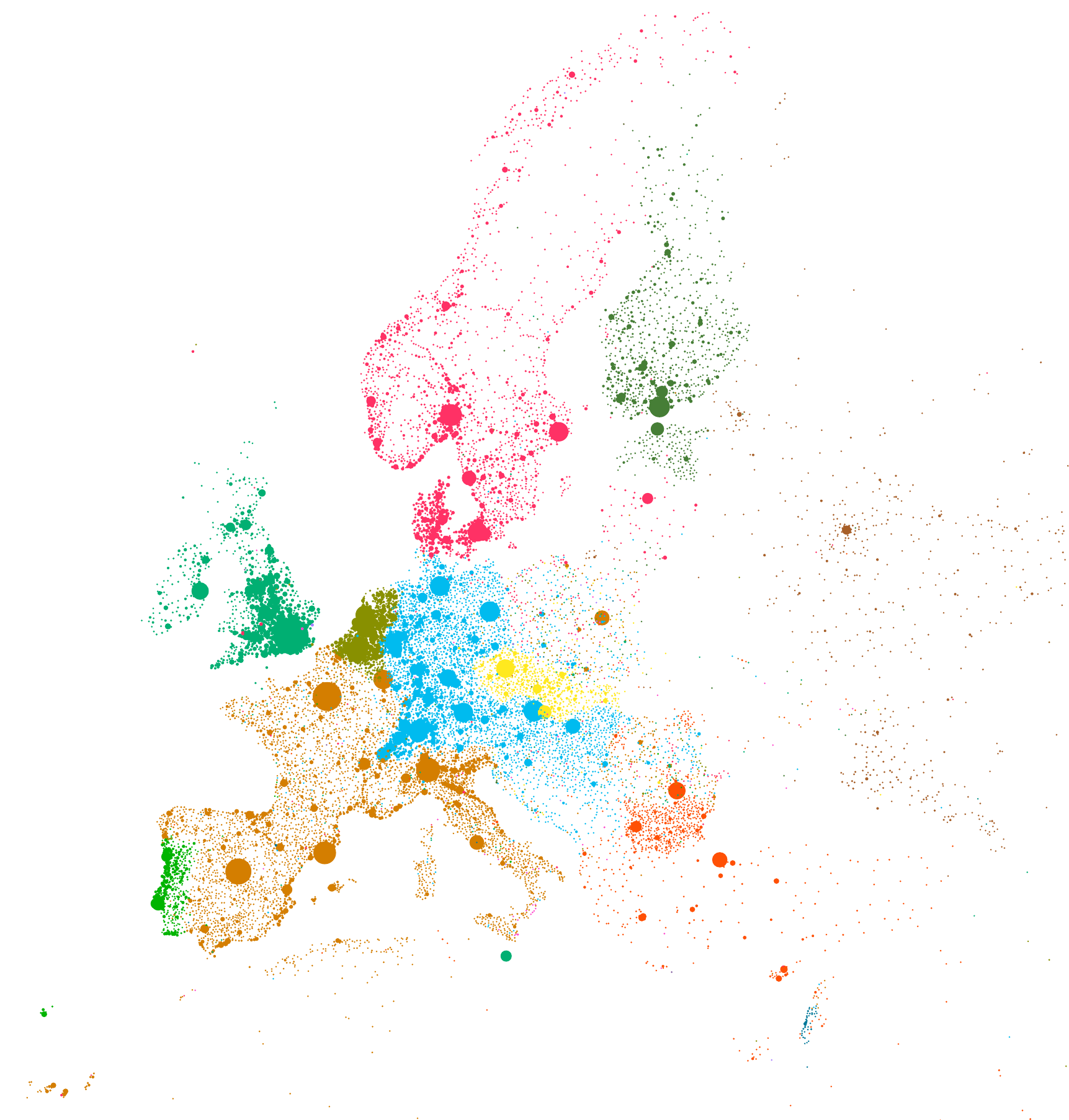} 
	\caption{Community structure of the European part of the network}
	\label{fig:europe}
\end{figure}

Figure~\ref{fig:global} gives a representation of the spatial distribution of the community structure. The nodes are cities, positioned in space based on their longitude and latitude. Cities in the same community have the same color. There are a number of observations. First of all, we see that this color-coding of the communities clearly highlights particular countries. Portugal is clearly separated from Spain on the Iberian Peninsula (but in the same community as Brazil), the United Kingdom (and Ireland) stands out, as do South Africa and Finland. Note that the community detection algorithm only considered board interlocks between firms in cities and did not have any information on nationality. Nonetheless, the results show that national borders do still play a role for the organization of business elites. Second, at the same time there are communities that bring together firms and cities from neighboring regions and countries. In Europe we see a southwestern and a central European community, and in Asia there are two communities as well. A closer look reveals that countries from the former British Empire are still relatively close and therefore grouped in one community: India, Australia and New Zealand. South Korea and Vietnam also belong to this group. 

Figure~\ref{fig:europe} zooms in on Europe and gives a more detailed insight in the spatial level of corporate elite organization. The European corporate network clearly separates in distinct regional communities, but a visual inspection already shows that national borders still play an important role. Interestingly, some of the communities reflect geo-political structures now long gone. One business community follows the boundaries of Czechoslovakia; another resembles the original early nineteenth century Kingdom of the Netherlands (the Netherlands, Belgium and Luxembourg). Among the Nordic countries, Finland separates from Scandinavia. Another interesting observation is that Poland is very divided. Poland seems to lack a cohesive business community as the German (or Central European) business community permeates Poland within the nineteenth century German and Prussian borders. These observations suggest that the level of cohesion among business elites differs significantly across Europe. 

\begin{figure}[!t]
	\includegraphics[width=0.8\textwidth]{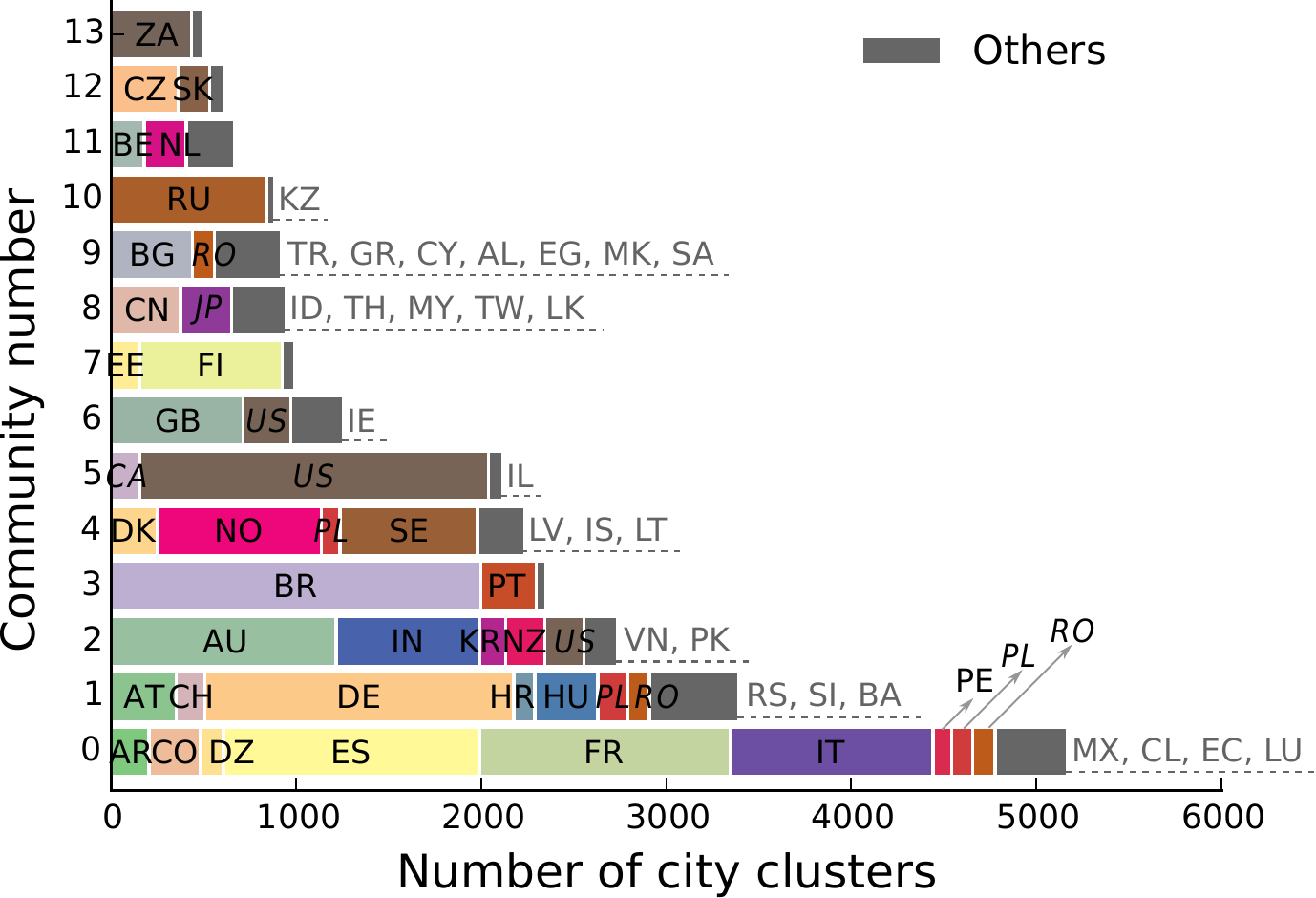} 
	\caption{National diversity of community composition}
	\label{fig:communities}
\end{figure}

For each community we can zoom in on the nationality of the cities that are part of the clusters (see Figure~\ref{fig:communities}). The smallest communities are the ones around South Africa, the Czech and Slovak republics, and The Netherlands and Belgium. The Russian corporate elite is organized in a separate community, and Russian cities are absent from the other communities (with only a handful exceptions). This is in line with the high betweenness centrality of Moscow, reported above. Moscow connects the distinct Russian community to the rest of the world.

Bulgarian and Romanian corporate elites dominate a South East European community, that extends south to regions in Turkey, Greece and Egypt. China and Japan together dominate the East Asian community, while Australia and India are the dominant countries in the more diverse and larger Asian community. This community also includes a number of North American as well as European cities. In Northern Europe, the Baltic countries are divided over the Scandinavian community (Latvia and Lithuania) and a Finnish one (Estonia). The latter reflects the deep cultural and language similarities. In a way the UK-USA cluster may reflect language and cultural similarities as well, although within the USA there is a division between locations oriented on the Anglo-American community; the Commonwealth Asian community, and a North American community. This North American community includes Israel as well, reflecting long standing political and social ties. Language similarity is also evident in the Brazilian-Portuguese community. Finally, there are two very large and internationally diverse communities in Europe. One is centered around Germany and extends south to the Alpine countries and east to Hungary, Croatia and parts of Poland. The largest community contains the business elites of the largest Southern European countries on the continent: Spain, France, and Italy. 

Reviewing the national composition of the communities, we need to conclude that there is considerable diversity. Some communities single out a particular business elite, such as the Russian one. Some bring together business elites from culturally similar regions, such as the transatlantic community; Scandinavia, and Portugal-Brazil. Other communities however are more diverse in its composition and suggest a level of corporate elite organization that goes across national borders.

\subsection{The nonlocal network ties}

\begin{figure}[!b]
	\includegraphics[width=\textwidth]{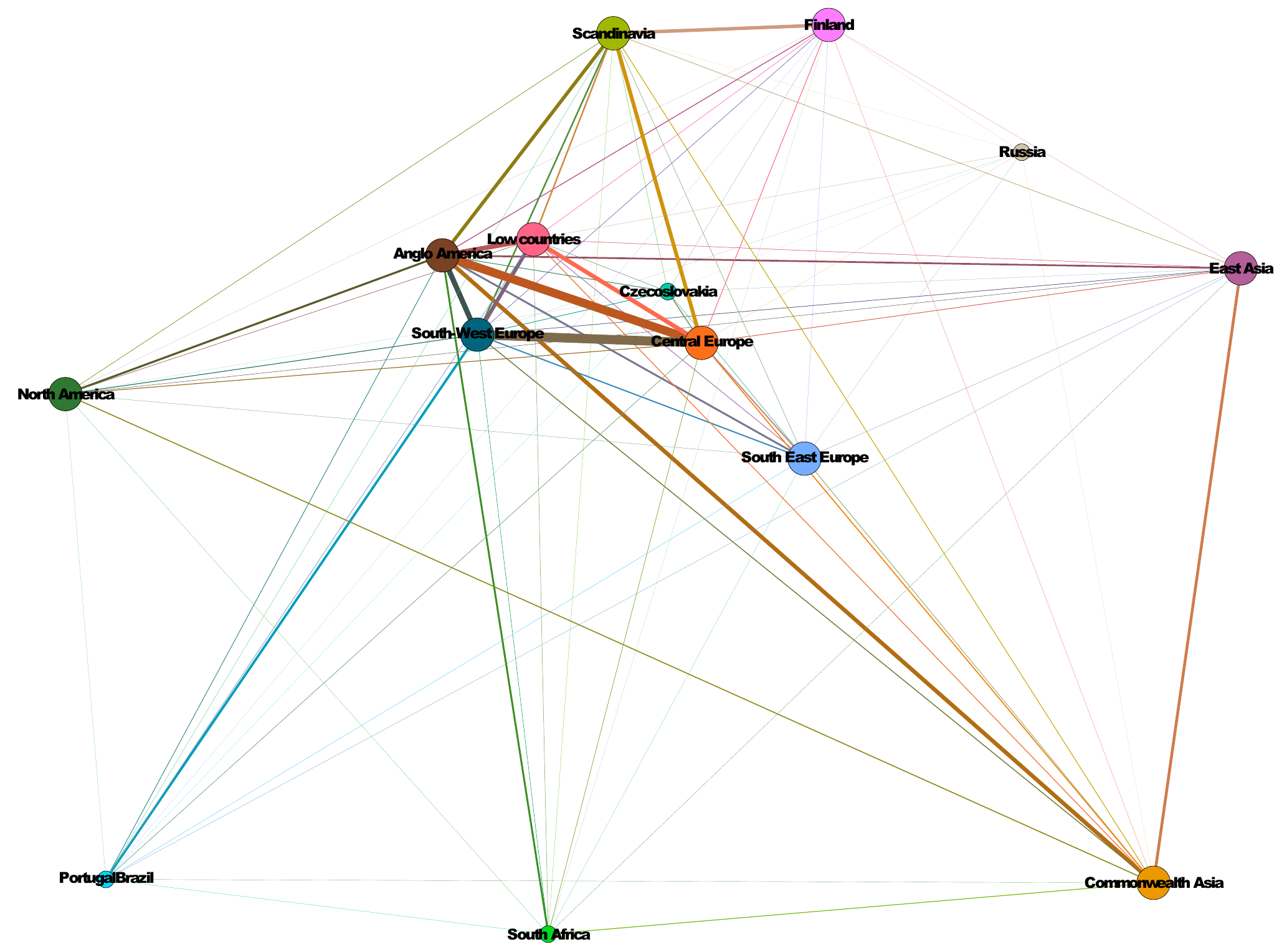} 
	\caption{Community by community network} 
	\label{fig:communitycommunity}
\end{figure}

The lion share of the board interlocks connect firms and cities that fall within the same community. If we only consider the binary network where cities are either connected or disconnected, 67 percent of all the connections are local and 33 percent nonlocal. If we count all board interlocks between cities (the weighted network), an impressive 92 percent of the network resides within the communities. Note that this is also reflected by the high modularity value. At the same time, the 8\% of nonlocal ties contain no less than 2.84 million board interlocks. These nonlocal interlocks connect the communities, and Figure~\ref{fig:communitycommunity} shows what the global network looks like if we aggregate to the level of communities. To ease the interpretation, the position of the nodes is related to their geographic position vis-à-vis each other. As we can expect, the network is fully connected. But some communities are better connected than others, as reflected by the thickness of the tie in Figure~\ref{fig:communitycommunity}. The strongest tie connects South West Europe with Central Europe. This is, indeed, the German-Franco axis that is so fundamental to the process of European integration. At the same time, there still is a clear separation between a French and German corporate sphere of influence. One may say that when the time comes that the German and French business elites are no longer distinct communities but part of the same community, the project of European Unification has come to the next level. Both the central and southwestern European communities are well connected with the Anglo-American community, with London as hub. The Commonwealth label of one of the communities seems justified given the strong tie it has with Anglo-America, in fact through the strong ties with London. 

\begin{figure}[hp]
	\includegraphics[angle=-90,width=0.92\textwidth]{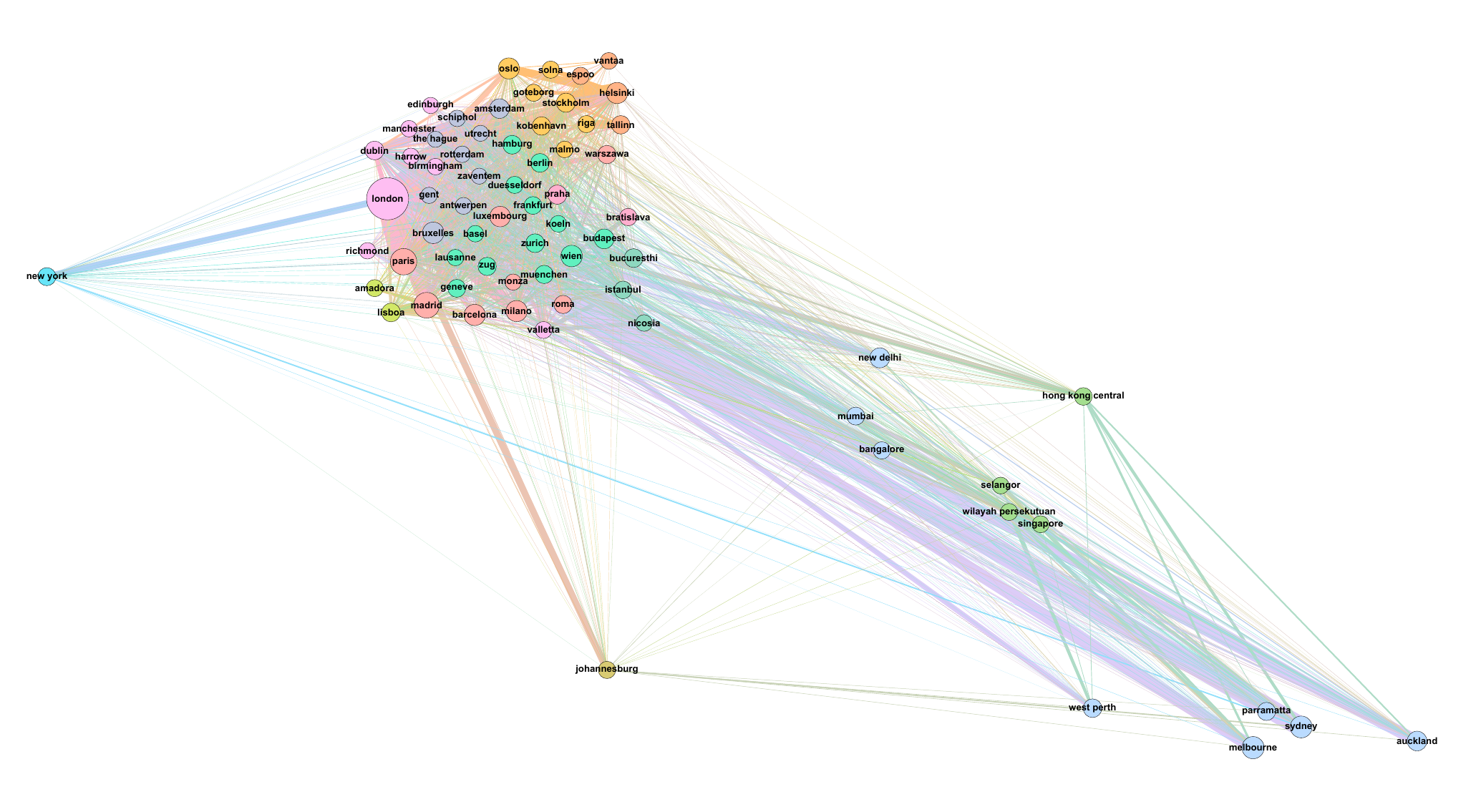} 
	\caption{Nonlocal network between the main networked cities} 
	\label{fig:nonlocal}
\end{figure}

	We already saw that London plays a key role in connecting business elites across the globe, and this remains the case when we consider the cities that are most important in spanning the global network of nonlocal corporate ties. Figure~\ref{fig:nonlocal} singles out the network between the most central cities in the network of nonlocal ties, that is the group of cities that connect to at least 1,000 cities in other communities. Given the relatively high density of ties in Europe and the larger number of European communities, it is to be expected to see a large number of European cities among this hard core. In Europe, the capital cities (Madrid, Paris, Berlin, Amsterdam, Brussels, Copenhagen and so on) are all part of the core as the network. We also see that major transportation hubs such as airports and harbors bring together the transnational corporate elite (Schiphol, Zaventem, Antwerpen, Rotterdam). Valetta, in south Europe, is the capital city of Malta. In Asia, it is noticeable that a number of Indian cities are part of the hard core of the network. Hong Kong singles out as a hub in Asia, reaching out to all other parts of the network. In Africa, Johannesburg remains well connected with a noticeable strong tie to London, as many other cities have as well. Interestingly, New York is the only USA city among this key group in the network in Figure~\ref{fig:nonlocal}. New York is the bridgehead of the USA to the rest of the world.

\section{Conclusion and discussion}

We argued that in order to advance our understanding of the organization of corporate elites, we must not force the  distinction between national and transnational upon the analysis. Business elites reconfigure their networks, from the city level, to the national level, and beyond. Here we asked what is the current level of organization. In order to develop an answer to this question we studied the worldwide network of interlocking directorates of over 5 million firms. We did so without assuming that the main distinction between corporate elites is that of the national -- transnational level. Building on the theoretical distinction between local and nonlocal ties we used network analysis and in particular community detection to understand how corporate elites are organized in distinct business communities, and where and how these business communities are subsequently interconnected. Our large-scale network analysis rendered a number of insights.

	First of all, national borders still play an important role for the organization of corporate elites. The community structure we uncovered shows that national borders still form cleavages for corporate elites. The community structure also echoes state delineations long gone, such as Czechoslovakia and the original Kingdom of the Netherlands, and also in part the mixture of communities in Poland that are partly structured along the old German border. Communities also reflect language and ethnic connections, for instance between Estonia and Finland or Portugal and Brazil. 
Second, we also find that what is local for the business elite is indeed changing from the national to the transnational plane. However, we see considerable heterogeneity in the level of elite organization across different regions in the world. In Russia, the local business community coincides with the state boundaries. In the USA in contrast we find three spheres of influence: one North American (including Israel) community, one community oriented towards Asia and one community oriented towards Europe, most prominently London. In this paper we only studied a snapshot image of the global network of interlocking directorates. A longitudinal study with similar setup could trace how these communities develop over time. In the case of the USA, this would render unique opportunities, for instance to study if the pivot to Asia in the USA foreign policy is also reflected in the orientation of the business elite. And, if it does, we can ask whether the business elite networks foreshadow the foreign policy change or follow its lead. Third, we found that the city hubs in the networks are very different in nature. London is the most central city in the network of the global corporate elite. The status of London within its community and the network as a whole is unmatched by any other city.


When analyzing trans-community or nonlocal networks, rather than transnational ones, we find that outside Europe, connections are primarily regional in nature and that the networks do not reach out much to the rest of the world. If the true `global corporate elites' are the ones that transcend their mostly regionally structured local business communities, then they can mainly be found in Europe or in the hubs that are well connected to London (often located in commonwealth territory).
Beside the relatively low representation of North American cities, cities in China, Japan and Russia do not reach out much outside their local communities at all. This sheds light on how the transnational or global elite is socially configured and whether we will indeed witness the emergence of a truly `global corporate elite' or `global capitalist class'. Despite claims on increasing interdependence of various economic regions in the world, corporate elites from Asia, Latin-America and even North-America, are not yet fully integrated in the corporate elite network that transcend local communities. 

In sum, it seems warranted to conclude that our approach to study local and nonlocal corporate elite ties is a fruitful endeavor. We hope that our work shows that it is important to include space in the study of corporate elites. In studies on corporate elites, space has long been implicitly considered as a natural stage upon which social action unfolds~\cite{friedland1994space}. 
Although there existed some notions of spatiality in studies of corporate elites (see for instance \cite{dooley1969interlocking,levine1972sphere,allen1978economic,mizruchi1982american,mintz1985power}), 
 geographic proximity of inter-firm relations was mostly regarded to be something unsocial. 
Following \cite{friedland1994space} and \cite {kono1998lost}, 
 we prefer to view space as a socially formed determinant, medium and outcome of social structure. Our approach shows that it is elementary, possible and helpful to study space without forcing a certain spatial category such as states upon the analysis. 
	
At the same time, we recognize that our study is limited in a number of ways. Here we considered board interlocks as a telltale sign of elite organization. In the introduction we already pointed out that we believe this is an expedient, but necessarily limited approach for studying corporate elites. For one, it leaves out key actors outside the boardroom such as lawyers, accountants, but also investors and fund managers who arguable have an increasing role in today’s financialized capitalism. We therefore do not want to claim that we have now investigated the corporate elite in its fullest extent. Another limitation is a direct result of the large scale of our analysis. Although the scale allows us to engage in a fine grained analysis of worldwide patterns of elite organization, it does so at the expense of a detailed understanding of its constituting individual components: the firms and directors. Our analysis is in that sense disconnected from the logics and interests that drive the formation of the networks that we study at the level of the actors out of which it is composed. And to make matters more complicated, we know that these logics and interests differ across different regions in the world. Board interlocks have a different meaning and function for firms in Asia, Latin America, or Europe. We do see very promising developments however in the area of modelling network dynamics (agent based and exponential random graph models) that may make it possible in the near future to comparatively study the generating mechanisms of board interlocks in the different communities that we uncovered. Finally, the large scale also means that we cannot be as confident about the quality of our data as we can when we manually collect samples. For instance, the low weight of the USA in the network may be due to lower data quality (\ref{sec:datamethods}). This is an important concern, and with the increase of large scale corporate network studies we must devote more attention towards the development of tools and techniques to gauge and improve data quality. 
  
An important benefit from our approach is that it allows for, and indeed uncovered, different levels of elite organization in different areas of the world. The community detection approach also allows for more fine grained and detailed studies, in at least two ways. First, we can continue and investigate the particularities of the network structure within communities. How are the different communities in their topological properties? And how does this relate to our theoretical understanding of elite formation in these regions? Second, as explained in the methods section, we can further dissect the community structure and investigate the hierarchically nested structure. We now have the data available, necessary to conduct such fine grained analysis. We see this as a promising way forward for understanding the multilevel structure of elite organization as pointed out in~\cite{heemskerk2016corporate}. 
This brings us to a final and related shortcoming of our study. We argue against the national--international dichotomy in the study of corporate elites. And while our approach breaks free from theoretical and methodological nationalism, our suggested dichotomy of local and nonlocal leaves only little room for a proper analysis of the multi-level structure we encountered. Follow up work must make an effort to develop a more multi-level or nested approach of social organization of corporate elites.

We believe our approach also presents a promising avenue for developing an answer to a key question in global politics, international relations, and economic sociology today: do we see a relatively cohesive transnational corporate power elite or are there in fact factions with irreconcilable differences among such a transnational corporate elite? The distinction between local and nonlocal corporate elite networks allows us to reinvigorate the classic debate on elite power between elitists and pluralists at the global plane. Elitist scholars from the power structure research school already pointed out that the question is mainly a matter of local and nonlocal elite organization. Both \cite{mills1956power} and \cite{domhoff1967rules} 
 noted that, whereas at the local level there was room for competition between elites as pointed out in \cite{price1962governs}, 
 the institutions that transcended the local level were dominated by a small number of people of a clearly discernible  upper class. This suggests a research agenda where we investigate in depth the characteristics, motives and strategies of the firms, directors, and perhaps also cities that build the nonlocal elite networks. 

\subsubsection*{Acknowledgements}
This research is part of the CORPNET project (see \url{http://corpnet.uva.nl}), which has received funding from the European Research Council (ERC) under the European Union’s Horizon 2020 research and innovation programme (grant agreement number 638946).

\bibliography{sociologica2016}

\begin{thebibliography}{40}
\providecommand{\natexlab}[1]{#1}
\providecommand{\url}[1]{\texttt{#1}}
\expandafter\ifx\csname urlstyle\endcsname\relax
  \providecommand{\doi}[1]{doi: #1}\else
  \providecommand{\doi}{doi: \begingroup \urlstyle{rm}\Url}\fi

\bibitem[Allen(1978)]{allen1978economic}
M.~P. Allen.
\newblock Economic interest groups and the corporate elite structure.
\newblock \emph{Social Science Quarterly}, 58\penalty0 (4):\penalty0 597--615,
  1978.

\bibitem[Blondel et~al.(2008)Blondel, Guillaume, Lambiotte, and
  Lefebvre]{blondel2008}
V.~Blondel, J.~Guillaume, R.~Lambiotte, and E.~Lefebvre.
\newblock Fast unfolding of communities in large networks.
\newblock \emph{Journal of Statistical Mechanics: Theory and Experiment},
  10:\penalty0 P10008, 2008.

\bibitem[Brandes(2008)]{brandes2008variants}
U.~Brandes.
\newblock On variants of shortest-path betweenness centrality and their generic
  computation.
\newblock \emph{Social Networks}, 30\penalty0 (2):\penalty0 136--145, 2008.

\bibitem[Burris and Staples(2012)]{burris2012search}
V.~Burris and C.~L. Staples.
\newblock In search of a transnational capitalist class: Alternative methods
  for comparing director interlocks within and between nations and regions.
\newblock \emph{International Journal of Comparative Sociology}, 53\penalty0
  (4):\penalty0 323--342, 2012.

\bibitem[Carroll and Fennema(2002)]{carroll2002there}
W.~K. Carroll and M.~Fennema.
\newblock Is there a transnational business community?
\newblock \emph{International Sociology}, 17\penalty0 (3):\penalty0 393--419,
  2002.

\bibitem[Carroll et~al.(2010{\natexlab{a}})Carroll, Carson, Fennema, Heemskerk,
  Sapinski, et~al.]{carroll2010making}
W.~K. Carroll, C.~Carson, M.~Fennema, E.~Heemskerk, J.~Sapinski, et~al.
\newblock \emph{The making of a transnational capitalist class: Corporate power
  in the twenty-first century}.
\newblock Zed books, 2010{\natexlab{a}}.

\bibitem[Carroll et~al.(2010{\natexlab{b}})Carroll, Fennema, and
  Heemskerk]{carroll2010constituting}
W.~K. Carroll, M.~Fennema, and E.~M. Heemskerk.
\newblock Constituting corporate europe: A study of elite social organization.
\newblock \emph{Antipode}, 42\penalty0 (4):\penalty0 811--843,
  2010{\natexlab{b}}.

\bibitem[Compston(2013)]{compston2013network}
H.~Compston.
\newblock The network of global corporate control: Implications for public
  policy.
\newblock \emph{Business and Politics}, 15\penalty0 (3):\penalty0 357--379,
  2013.

\bibitem[David and Westerhuis(2014)]{david2014power}
T.~David and G.~Westerhuis.
\newblock \emph{The power of corporate networks: A comparative and historical
  perspective}.
\newblock Routledge, 2014.

\bibitem[Davis and Greve(1997)]{davis1997corporate}
G.~F. Davis and H.~R. Greve.
\newblock Corporate elite networks and governance changes in the 1980s.
\newblock \emph{American Journal of Sociology}, 103\penalty0 (1):\penalty0
  1--37, 1997.

\bibitem[Domhoff(1967)]{domhoff1967rules}
G.~W. Domhoff.
\newblock \emph{Who Rules America?}
\newblock Prentice-Hall, 1967.

\bibitem[Domhoff(1970)]{domhoff1970}
G.~W. Domhoff.
\newblock \emph{Higher circles: The governing class in America}.
\newblock Random House, 1970.

\bibitem[Dooley(1969)]{dooley1969interlocking}
P.~C. Dooley.
\newblock The interlocking directorate.
\newblock \emph{The American Economic Review}, 59\penalty0 (3):\penalty0
  314--323, 1969.

\bibitem[Fennema(1982)]{fennema1982international}
M.~Fennema.
\newblock \emph{International networks of banks and industry}.
\newblock Martinus Nijhoff Publishers, 1982.

\bibitem[Fennema and Schijf(1985)]{fennema1985transnational}
M.~Fennema and H.~Schijf.
\newblock The transnational network.
\newblock In \emph{Networks of corporate power: A comparative analysis of ten
  countries}, pages 250--266. Polity Press, 1985.

\bibitem[Friedland and Palmer(1994)]{friedland1994space}
R.~Friedland and D.~Palmer.
\newblock Space, corporation and class: Toward a grounded theory.
\newblock In \emph{NowHere: Space Time and Modernity.}, pages 287--334.
  University of California Press, 1994.

\bibitem[Fukunaga and Hostetler(1975)]{fukunaga1975estimation}
K.~Fukunaga and L.~D. Hostetler.
\newblock The estimation of the gradient of a density function, with
  applications in pattern recognition.
\newblock \emph{IEEE Transactions on Information Theory}, 21\penalty0
  (1):\penalty0 32--40, 1975.

\bibitem[Heemskerk(2007)]{heemskerk2007decline}
E.~M. Heemskerk.
\newblock \emph{Decline of the corporate community: Network dynamics of the
  Dutch business elite}.
\newblock Amsterdam University Press, 2007.

\bibitem[Heemskerk and Takes(2016)]{heemskerk2016corporate}
E.~M. Heemskerk and F.~W. Takes.
\newblock The corporate elite community structure of global capitalism.
\newblock \emph{New Political Economy}, 21\penalty0 (1):\penalty0 90--118,
  2016.

\bibitem[Heemskerk et~al.(2013)Heemskerk, Daolio, and
  Tomassini]{heemskerk2013community}
E.~M. Heemskerk, F.~Daolio, and M.~Tomassini.
\newblock The community structure of the european network of interlocking
  directorates 2005--2010.
\newblock \emph{PloS one}, 8\penalty0 (7):\penalty0 e68581, 2013.

\bibitem[Heemskerk et~al.(2016)Heemskerk, Fennema, and
  Carroll]{heemskerk2016global}
E.~M. Heemskerk, M.~Fennema, and W.~K. Carroll.
\newblock The global corporate elite after the financial crisis: Evidence from
  the transnational network of interlocking directorates.
\newblock \emph{Global Networks}, 16\penalty0 (1):\penalty0 68--88, 2016.

\bibitem[Kentor and Jang(2004)]{kentor2004yes}
J.~Kentor and Y.~S. Jang.
\newblock Yes, there is a (growing) transnational business community a study of
  global interlocking directorates 1983--98.
\newblock \emph{International Sociology}, 19\penalty0 (3):\penalty0 355--368,
  2004.

\bibitem[Kleinberg(2000)]{kleinberg2000small}
J.~Kleinberg.
\newblock The small-world phenomenon: An algorithmic perspective.
\newblock In \emph{Proceedings of the 32nd Annual ACM Symposium on Theory of
  Computing (STOC)}, pages 163--170, 2000.

\bibitem[Kono et~al.(1998)Kono, Palmer, Friedland, and Zafonte]{kono1998lost}
C.~Kono, D.~Palmer, R.~Friedland, and M.~Zafonte.
\newblock Lost in space: The geography of corporate interlocking directorates.
\newblock \emph{American Journal of Sociology}, 103\penalty0 (4):\penalty0
  863--911, 1998.

\bibitem[Levine(1972)]{levine1972sphere}
J.~H. Levine.
\newblock The sphere of influence.
\newblock \emph{American Sociological Review}, 37\penalty0 (1):\penalty0
  14--27, 1972.

\bibitem[Mills(1956)]{mills1956power}
C.~W. Mills.
\newblock \emph{The power elite}.
\newblock Oxford University Press, 1956.

\bibitem[Mintz and Schwartz(1985)]{mintz1985power}
B.~A. Mintz and M.~Schwartz.
\newblock \emph{The power structure of American business}.
\newblock University of Chicago Press, 1985.

\bibitem[Mizruchi(1982)]{mizruchi1982american}
M.~S. Mizruchi.
\newblock \emph{The American corporate network, 1904--1974}.
\newblock Sage, 1982.

\bibitem[Mizruchi(1996)]{mizruchi1996interlocks}
M.~S. Mizruchi.
\newblock What do interlocks do? an analysis, critique, and assessment of
  research on interlocking directorates.
\newblock \emph{Annual Review of Sociology}, pages 271--298, 1996.

\bibitem[Mizruchi(2013)]{mizruchi2013fracturing}
M.~S. Mizruchi.
\newblock \emph{The fracturing of the American corporate elite}.
\newblock Harvard University Press, 2013.

\bibitem[Molotch(1976)]{molotch1976city}
H.~Molotch.
\newblock The city as a growth machine: Toward a political economy of place.
\newblock \emph{American Journal of Sociology}, pages 309--332, 1976.

\bibitem[Pak(2013)]{pak2013gentlemen}
S.~J. Pak.
\newblock \emph{Gentlemen bankers}.
\newblock Harvard University Press, 2013.

\bibitem[Price and Dahl(1962)]{price1962governs}
H.~D. Price and R.~A. Dahl.
\newblock \emph{Who Governs? Democracy and power in an American city}.
\newblock Yale University Press, 1962.

\bibitem[Robinson(2004)]{robinson2004theory}
W.~I. Robinson.
\newblock \emph{A theory of global capitalism: Production, class, and state in
  a transnational world}.
\newblock Johns Hopkins University Press, 2004.

\bibitem[Scott(1991)]{scott1991networks}
J.~Scott.
\newblock Networks of corporate power: A comparative assessment.
\newblock \emph{Annual Review of Sociology}, pages 181--203, 1991.

\bibitem[Sklair(2001)]{sklair2001transnational}
L.~Sklair.
\newblock \emph{The transnational capitalist class}.
\newblock Blackwell Publishers, 2001.

\bibitem[Takes and Kosters(2013)]{takes2013computing}
F.~W. Takes and W.~A. Kosters.
\newblock Computing the eccentricity distribution of large graphs.
\newblock \emph{Algorithms}, 6\penalty0 (1):\penalty0 100--118, 2013.

\bibitem[Useem(1984)]{useem1984inner}
M.~Useem.
\newblock \emph{The inner circle}.
\newblock Oxford University Press, 1984.

\bibitem[Vitali et~al.(2011)Vitali, Glattfelder, and
  Battiston]{vitali2011network}
S.~Vitali, J.~B. Glattfelder, and S.~Battiston.
\newblock The network of global corporate control.
\newblock \emph{PloS one}, 6\penalty0 (10):\penalty0 e25995, 2011.

\bibitem[Zeitlin(1974)]{zeitlin1974orporate}
M.~Zeitlin.
\newblock Corporate ownership and control: The large corporation and the
  capitalist class.
\newblock \emph{American Journal of Sociology}, 79\penalty0 (5):\penalty0
  1073--1119, 1974.

\end{thebibliography}

\newpage
\appendix

\section{City level aggregation}
\label{sec:cityclustermethod}

As explained in Section~\ref{sec:datamethods}, we study the network of interlocking directorates at the level of cities. 
However, if we simply merge firms from the same city into one node in order to create a city-by-city network, we run into a number of problems due to the noise and inconsistencies in the city information. 
For example, we would have to deal with spelling variations of city names, e.g., The Hague, the political capital of the Netherlands, is sometimes written as 's-Gravenhage. 
And for example Brussels, the capital of Belgium, is also written as Bruxelles (French) and sometimes as Brussel (Dutch). 
More importantly, small suburban areas close to a larger city may have a different name, but would by any expert be seen as belonging to the same metropolitan area. 
However, due to the sheer amount of information we are not able to manually correct this. 

The aforementioned problems are solved by retrieving for each city its coordinates (latitude and longitude) using the Google Maps API, and then grouping cities with very similar coordinates into one node which we call a citycluster. 
We used the well-known and established MeanShift algorithm~\citep{fukunaga1975estimation} as implemented in the \texttt{sklearn} package in Python to cluster the cities. 
For one out of ten of the cities, we only had information on the country in which the firm is located, or we were not able to automatically determine the latitude and longitude, leaving us no choice but to disregard these firms in our study. 
The majority (90\%) of these firms with missing location data were however located in Panama, leaving only 1\% missing data for the remainder of the world. 
Typically, these firms were smaller businesses in rural areas, having mostly national interlocks, if any. 
These steps result in a citycluster-by-citycluster network in which cityclusters are connected by a weighted edge denoting the number of firms that are connected between two clusters. 
In some cityclusters we find cities from a border region of two different countries. 
In these instances, we split the original citycluster into two cityclusters, each containing the cities on one side of the border. 
In the text, we simply refer to cities instead of cityclusters.

\end{document}